\documentstyle[12pt,amsmath,amssymb,graphicx,mathtools,verbatim,bbold,appendix]{article}

\def\be{\begin{eqnarray}}
\def\ee{\end{eqnarray}}
\def\ba{\begin{array}}
\def\ea{\end{array}}
\def\nn{\nonumber}
\def\l{\langle}
\def\r{\rangle}
\def\p{\partial}

\def\Tr{{\rm Tr}\,}
\def\e{\epsilon}
\makeatletter
\def\Appendix{\appendix
  \def\@seccntformat##1{Appendix~\csname the##1\endcsname.~~}}
\makeatother
\DeclareSymbolFont{extraup}{U}{zavm}{m}{n}
\DeclareMathSymbol{\varheart}{\mathalpha}{extraup}{86}
\DeclareMathSymbol{\vardiamond}{\mathalpha}{extraup}{87}

\hoffset= - 0.9in         % switch off for draft style
\title{{\bf  $N=1$ superconformal blocks with Ramond fields \newline from AGT correspondence}
\vspace{.2cm}}

\begin{document}

\author{Alexander Belavin$^{1,2,3}$, Baur Mukhametzhanov$^{1,2}$\footnote{E-mail:  baur@itp.ac.ru} 
\vspace*{10pt}\\[\medskipamount]
$^1$~\parbox[t]{0.88\textwidth}{\normalsize\it\raggedright
L.D.Landau Institute for Theoretical Physics,
142432 Chernogolovka, Russia}
\vspace*{10pt}\\[\medskipamount]
$^2$~\parbox[t]{0.88\textwidth}{\normalsize\it\raggedright
Moscow Institute of Physics and Technology, 141700 Dolgoprudny, Russia}
\vspace*{10pt}\\[\medskipamount]
$^3$~\parbox[t]{0.88\textwidth}{\normalsize\it\raggedright
Institute for Information Transmission Problems, 127994, Moscow, Russia}}

\baselineskip 0.6cm

\date{}
\maketitle
 \vspace{-0.9cm}
%\magnification=1200

\vspace{1cm}

\begin{abstract}
We use AGT correspondence between  $N=2$ SUSY Yang-Mills theory on ${\mathbb R}^4/{\mathbb Z}_2$ and two-dimensional CFT model with the algebra $ {\cal H} \oplus \hat{sl}(2)_2 \oplus \text{NSR}$ to obtain the explicit expressions 
for 4-point NSR conformal blocks including Ramond fields in terms of Nekrasov partition functions and
correlation functions of $\hat{sl}(2)_2$ WZW model. 
\end{abstract}

 \tableofcontents

\section{Introduction}

The remarkable relation between $N=2$ SUSY Yang-Mills theories and 2d CFT, discovered in \cite{Alday:2009aq}, 
provides a possibility for writing explicit combinatorial expressions for conformal blocks of different 
2d CFTs \cite{Wyllard:2009hg, Mironov:2009by, Alday:2010vg, Belavin:2011pp, Belavin:2011tb, Bonelli:2011jx, 
Bonelli:2011kv, Wyllard:2011mn, Alfimov:2011ju, Belavin:2012uf, Belavin:2012qh, Ito:2011mw} in terms of 
Nekrasov partition functions \cite{Nekrasov:2002qd}.

We believe that the basic reason for AGT correspondence is a possibility for some geometrical realization of 
representations of infinite dimensional conformal algebras on the spaces of the equivariant cohomologies of 
Instanton moduli spaces  ${\cal M}$. Historically, the first examples of such constructions have been given 
by Nakajima \cite{Nakajima1997, Nakajima1998} for cases of Heisenberg or Kac-Moody algebras  .     

There exists  a natural distinguished basis  in the space of the equivariant cohomologies. The vectors of this 
basis are labelled by the fixed points of the maximal abelian subgroup (torus group), which acts on the instanton 
moduli space \cite{Atiyah}. This basis is orthogonal and norms of the basis vectors equal to the determinants of the 
vector field of the torus group at the fixed points. Matrix elements of geometrically defined vertex operators in this basis have 
completely factorized form and coincide with Nekrasov partition function for the bifundamental representation. 

For $SU(2)$ gauge theory  on ${\mathbb R}^4$ this basis was constructed in \cite{Alba:2010qc}. 
In fact this construction has proved the  AGT conjecture for this case.   

It was suggested in \cite{Belavin:2011pp} that $N=2$ SUSY $SU(r)$ gauge theory on ${\mathbb R}^4/{\mathbb Z}_p$ 
is in AGT correspondence with the coset conformal field theory with the algebra

\be 
{\cal A}(r,p) = {\hat{gl}(n)_r \over \hat{gl}(n-p)_r}
\ee 
which can be rewritten using the level-rank duality as 

\be  
{\cal A}(r,p) = {\cal H} \oplus \hat{sl}(p)_r \oplus \frac{\hat{sl}(r)_p \oplus \hat{sl}(r)_{n-p}  }{\hat{sl}(r)_n}
\ee 

Parameter $n$ in this algebra turns out to correspond to the Nekrasov's $\Omega$-background parameters $\e_1,\e_2$. Since expressions for $SU(r)$ partition functions on ${\mathbb R}^4/{\mathbb Z}_p$ are known \cite{Fucito:2004ry}, this correspondence allows us to write explicit expressions for correlation functions in CFT models with the algebra 
${\cal A}(r,p)$. 

Since we anticipate that the torus fixed points of the moduli space correspond to the basis vectors in the representation of the CFT algebra ${\cal A}(r,p)$, we can make a preliminary check of the conjecture  \cite{Belavin:2011pp}. Namely we can compare the generating functions for the number of the fixed points of the torus 
with the characters of representations of ${\cal A}(r,p)$. This kind of checks are already non-trivial support of the conjecture. 

These ideas were partially realized in \cite{Belavin:2011pp, Belavin:2011tb, Belavin:2011sw} for the case of $SU(2)$ instantons on ${\mathbb R}^4/{\mathbb Z}_2$ and corresponding algebra ${\cal A}(2,2) = {\cal H} \oplus \hat{sl}(2)_2 \oplus \text{NSR}$ in Neveu-Schwarz sector. Also the 4-point conformal blocks of the type $\l \Phi R R \Phi\r$  were treated in Whittaker limits in \cite{Ito:2011mw}, where $\Phi$ is Neveu-Schwarz field and $R$ is Ramond field.   

In this work we continue studying an important case of the correspondence between $SU(2)$ gauge theory on ${\mathbb R}^4/{\mathbb Z}_2$ and CFT with the algebra ${\cal A}(2,2) = {\cal H} \oplus \hat{sl}(2)_2 \oplus \text{NSR}$. We consider conformal blocks with Ramond fields of the type $\l \Phi R R \Phi\r$ and $\l RRRR \r$.

In the second section we discuss the moduli space of instantons, different pieces of the algebra ${\cal A}(2,2)$ and make a comparison of the characters of representations of this algebra with generating functions of the pairs of Young diagrams coloured in two colors. In the third section we introduce conformal blocks and equations which define them. And finally in the fourth section we make a general conjecture for the relation which is verified on 
the first two levels in appendix A,B.

\section{ Instantons of $SU(2)$ Yang-Mills theory on ${\mathbb R}^4/{\mathbb Z}_2$ \\
and algebra ${\cal A}(2,2)  = {\cal H} \oplus \hat{sl}(2)_2 \oplus \text{NSR}$}
\subsection{Moduli space and torus fixed points}

Let us denote by ${\cal M}(r,N)$ a smooth compactified moduli space of $SU(r)$ instantons on ${\mathbb R}^4$ with the topological number $N$. Then  the moduli space of Instantons of $SU(2)$ Yang-Mills theory on ${\mathbb R}^4/{\mathbb Z}_2$  is ${\cal M}(2,N)^{{\mathbb Z}_2}$ ~-- ${\mathbb Z}_2$-invariant part of the moduli space ${\cal M}(2,N)$. 
 
Our main  working conjecture, which we check and then use, looks as 

\be 
\label{general conj}
\boxed{
\sqcup_N {\cal M}(2,N)^{{\mathbb Z}_2} \Longleftrightarrow  {\cal H} \oplus \hat{sl}(2)_2 \oplus \text{NSR}
}
\ee
This relation should be understood as follows. The conformal algebra $ {\cal H} \oplus \hat{sl}(2)_2 \oplus \text{NSR}$ acts on equivariant cohomologies of $\sqcup_N {\cal M}(2,N)^{{\mathbb Z}_2}$. The instanton partition function calculated as an integral over this moduli space coincides with the conformal blocks in the CFT model ${\cal H} \oplus \hat{sl}(2)_2 \oplus \text{NSR}$.

The torus action on ${\cal M}(2,N)^{{\mathbb Z}_2}$ is induced by the action of an abelian subgroup of 
Lorentz group on ${\mathbb R}^4$ and $U(1)$ global transformations of the gauge group on framing at infinity. 
The fixed points of this torus are labelled by pairs $(Y_1, Y_2)$ of Young diagrams coloured in 2 colors \cite{Poghossian2011}. 
Also to each diagram we assign a charge $0$ or $1$ which is just the color 
of the box in the corner of the diagram. 

To perform  the  check of our conjecture   we will compare  the generating functions for the number of the    of pairs of Young diagrams with  charges $(\sigma_1,\sigma_2)=(0,0)$ or $(1,1)$ and  charges $(\sigma_1,\sigma_2)=(1,0)$ or $(0,1)$ against  characters of NS-representation and R-representation of algebra ${\cal A}(2,2)$.

Let us calculate the generating functions for the number of pairs of Young diagrams.
To do this calculation we need the character $\chi^{(1)}_{k,\sigma}(q)$ for the number of single diagrams with charge $\sigma$ and difference between white and black cells $k$ (see \cite{Belavin:2011tb})

\be
\chi^{(1)}_{k,\sigma}(q) = 
\begin{cases}
q^{k^2-{k\over 2}} \chi_B^2(q), \quad \text{for $\sigma = 0$} \\
q^{k^2+{k\over 2}} \chi_B^2(q), \quad \text{for $\sigma = 1$}
\end{cases}
\ee
where  $\chi_B(q) = \prod_{n=1}^{\infty}{1\over 1-q^n}$ - is the bosonic character.
Then one defines the character for the number of pairs of diagrams with charges $(\sigma_1,\sigma_2)$

\begin{align}
\chi^{(2)}_{\sigma_1,\sigma_2}(q,t) = \sum_{Y_1^{\sigma_1}, Y_2^{\sigma_2}} q^{|\vec{Y}| \over 2} t^{{(\sigma_1+\sigma_2) \over 2} +N_{+}-N_{-}} =
\sum_{k_1,k_2 \in {\mathbb Z}} \chi^{(1)}_{k_1,\sigma_1}(q) \chi^{(1)}_{k_2,\sigma_2}(q) t^{{(\sigma_1+\sigma_2) \over 2}+k_1+k_2} 
\end{align}
where $N_{\pm}$ ~--- is the number of white and black cells.

Thus the characters in NS-sector is

\begin{align}
\chi_{NS}(q,t) &= \chi^{(2)}_{0,0}(q,t) + \chi^{(2)}_{1,1}(q,t) = \chi_B^4(q) \sum_{ \substack{k_1,k_2 \in {\mathbb Z} \text{  or} \\ k_1,k_2 \in {\mathbb Z}+{1\over 2}}} q^{2k_1^2+2k_2^2-k_1-k_2 \over 2}t^{k_1+k_2} = \nn \\
\label{diagram char NS}
&= \chi_B^4(q) \sum_{k \in {\mathbb Z}}q^{k^2 \over 2} \sum_{n \in {\mathbb Z}}q^{n^2-n \over 2}t^n 
= 
\chi^2_{{\cal F},NS}(q)\chi^3_B(q)  \sum_{n \in {\mathbb Z}}q^{n(n-1) \over 2}t^{n} 
\end{align}
where $\chi_{{\cal F},NS} = \prod_{n=1}^{\infty}(1+q^{n-{1\over 2}})$ ~--- is the half-integer fermion character. 
Here we shifted the summation index in $\chi^{(2)}_{1,1}(q,t)$ by half in the second equality, then changed indices $k_1,k_2$ to $n=k_1+k_2, k=2k_1-n$ in the third equality and finally used Jacobi triple identity.

The character in R-sector is
\begin{align}
\chi_{R}(q,t)  &=  \chi^{(2)}_{1,0}(q,t) + \chi^{(2)}_{0,1}(q,t) = \chi_B^4(q) \sum_{ \substack{k_1,k_2 \in {\mathbb Z}  \text{  or}  \\ k_1,k_2 \in {\mathbb Z}+{1\over 2}}} q^{2k_1^2+2k_2^2-k_1+k_2 \over 2}t^{k_1+k_2+{1\over 2}} = \nn \\
&=\chi_B^4(q)  \sum_{k \in {\mathbb Z}}q^{(k-{1\over 2})^2 \over 2} \sum_{n \in {\mathbb Z}}q^{(n^2-{1\over 4}) \over 2}t^{n+{1\over 2}} =
2 q^{1\over 8}\chi^2_{{\cal F},R}(q) \chi^3_B(q) \sum_{n \in {\mathbb Z}+{1\over 2}}q^{n(n-1) \over 2}t^{n}
\label{diagram char R}
\end{align}
where $\chi_{{\cal F},R} = \prod_{n=1}^{\infty}(1+q^{n})$ ~--- is the integer fermion character. Here we shifted index of summation in $\chi^{(2)}_{0,1}(q,t)$ as $k_1 \to k_1+{1\over 2}, k_2 \to k_2-{1\over 2}$ in the second equality, then changed indices $k_1,k_2$ to $n=k_1+k_2, k=2k_1-n$ in the third equality and again used Jacobi triple identity.

\subsection{Algebra ${\cal A}(2,2)$}

Let us consider now the pieces of the algebra ${\cal A}(2,2) = {\cal H}\oplus \hat{sl}(2)_2 \oplus \text{NSR}$. 

First of all the Heisenberg algebra ${\cal H}$ has commutation relations  
$[a_n,a_m]= n \delta_{n+m}$ and the character of the Heisenberg algebra is simply a bosonic character $\chi_B(q)$.

The next one is NSR algebra. It has generators $L_n,G_r$ with commutation relations

\begin{align} 
\label{NSR algebra LL}
  [L_n,L_m] &= (n-m)L_{n+m}+{c\over 8}(n^3-n)\delta_{n+m} \\
\{G_r,G_s\} &= 2L_{r+s}+{c\over 2}(n^2-{1\over 4})\delta_{r+s} \\
\label{NSR algebra LG}
  [L_n,G_r] &= ({n\over 2}-r)G_{n+r}
\end{align}
where $n,m$~-- are integer and $r,s$~-- are integer in Ramond-sector and half-integer in NS-sector.
Representations of this algebra are constructed as following

\begin{align}
& L_n |\Delta\r = G_r |\Delta\r = 0, \quad \text{for $n,r>0$} \\
& L_0 |\Delta\r = \Delta |\Delta\r \\
& \pi = \{ L_{-n_1}\dots L_{-n_s}G_{-r_1}\dots G_{-r_l}|\Delta\r | n_1 \geq \dots \geq n_s, r_1 > \dots > r_l  \}
\end{align}
We denote them as $\pi_{NS}$~-- in NS-sector and $\pi_R$~-- in R-sector.
These representations have characters in NS-sector and R-sector

\be 
\chi_{\text{NS}}(q) = \chi_B(q) \chi_{{\cal F},NS}(q); \quad \chi_{\text{R}}(q)= \chi_B(q) \chi_{{\cal F},R}(q)
\ee

The last piece of ${\cal A}(2,2)$ we need to consider is $\hat{sl}(2)_k$ current algebra. 
It has generators $J_n^{a}$ which satisfy the commutation relations

\be
[J_n^a,J_m^b] = \e^{abc}J_{n+m}^c + {k\over 2} \delta_{n+m}\delta_{a,b}, \quad a,b,c=x,y,z
\ee

These commutation relations are agreed with OPEs of $J^\pm = J^x \pm J^y$ and $J^z$ 

\begin{align}
\label{JJ1}
J^z(u)J^z(v) &= {k/2 \over (u-v)^2} \\
J^z(u)J^{\pm}(v) &= {\pm J^{\pm}(v) \over u-v} \\
J^{+}(u)J^{-}(v) &= {k\over (u-v)^2} + {2J^z(v) \over u-v}
\label{JJ3}
\end{align}

Representations $\pi_j$ of $\hat{sl}(2)_k$ are constructed from the highest weight state $|j\r$

\begin{align}
&J_n^a |j\r = J_0^+ |j\r =0, \quad n>0, \\
&J_0^z |j\r = j |j\r \\
&\pi_j  = \{ J_{-n_1}^{a_1}\dots J_{-n_s}^{a_s} (J_{0}^{-})^t |j\r | n_1 \geq \ldots \geq n_s > 0, t\geq 0 \}
\end{align}

We are only interested in the so called integrable representations \cite{Kac},\cite{Goddard:1986ee} 
when the spin takes values: $j = 0,{1\over 2},1,\dots,(k-1) $. Thus in our case $k=2$ and $j=0,{1\over 2},1$. 
States $|j\r$ are known to have weights under the action of $L_0$: $\Delta_j = {j(j+1) \over k+2}$. 
Thus for $\hat{sl}(2)_2$ algebra there are three states with weights
\be
\label{sl(2) weights}
\Delta_{j=0} = 0, \quad \Delta_{j={1\over 2}} = {3\over 16}, \quad  \Delta_{j=1} = {1\over 2} 
\ee

Current algebra $\hat{sl}(2)_2$ has a realization in terms of free fermion and boson fields \cite{Fateev:1985mm}

\begin{align}
\label{free fields rep for currents1}
J^{\pm} &= \sqrt{2} \psi e^{\pm \phi} \\
J^z &=  \p \phi
\label{free fields rep for currents2}
\end{align}
OPEs (\ref{JJ1})~-- (\ref{JJ3}) are satisfied for $k=2$ if we take the OPEs of free fields as 

\be
\psi(z)\psi(0) \sim {1\over z}, \quad \phi(z)\phi(0) \sim \log z
\ee

One can also construct representations of free boson and fermion algebra in terms of their Laurent modes

\begin{align}
\label{free Laurent}
\psi(z)=\sum_{r} {\psi_r \over z^{r+{1\over 2}}}, \quad \phi(z) = Q+a_0 \log z - \sum_{n \neq 0} {a_n \over n z^n}
\end{align}
where the non-zero commutation relations are: $\{\psi_r,\psi_s \} = \delta_{r+s}, [Q,a_0]=1$ and $a_n$~-- satisfy Heisenberg commutation relations. 
There exists two types of representations of this free fields algebra: Neveu-Schwarz(NS) representation $F_{NS}^p$, 
$r \in {\mathbb Z}+{1\over 2}$ and Ramond(R) representation $F_R^p$, $r \in {\mathbb Z}$. These representations are constructed as following

\begin{align}
& \psi_r |p\r = a_n |p\r = 0,\quad n>0,r > 0 \\
\label{a_0 eigenvalue}
& a_0 |p\r = p |p\r \\
& F_{NS}^p = 
\{ \dots \psi_{-r_2}\psi_{-r_1} \dots a_{-n_2}a_{-n_1} |p\r | 0<n_1 \leq n_2 \dots , 0<r_1 < r_2< \dots ; r_1,r_2\ldots \in {\mathbb Z}+{1\over 2} \}  \label{free field rep} \\
& F_{R}^p = \{  \dots \psi_{-m_2} \psi_{-m_1}\dots a_{-n_2}a_{-n_1} |p\r | 0<n_1 \leq n_2 \dots , 0 \leq m_1<m_2<\ldots ; m_1,m_2,\dots \in {\mathbb Z} \} 
\end{align}

In this terms the representations of current algebra $\hat{sl}(2)_2$ look as

\begin{align}
\pi_0 \oplus \pi_1 &= \oplus_{n\in {\mathbb Z}} F_{NS}^n  \label{rep relation NS}\\
\pi_{1\over 2} &= \oplus_{n \in {\mathbb Z}+{1\over 2}} F^{n}_R \label{rep relation R}
\end{align}

Following the work \cite{Belavin:2011tb} we assume that  the fields which take part in AGT relation  are in  the following representations of the algebra ${\cal A}(2,2)$
\be 
\pi_{\cal H} \otimes \left(\pi_0 \oplus \pi_1\right) \otimes \pi_{\text{NS}} \quad \text{or} \quad \pi_{\cal H} \otimes \pi_{1\over 2} \otimes \pi_{\text{R}}  
\ee

Therefore in the next sections we will find the characters of these representations to compare them with 
the corresponding generation functions of the colored Young diagrams. 

\subsubsection{Character of ${\cal A}(2,2)$ representation in NS sector}

We will show in this subsection that the character of the representation 
$\pi_{\cal H} \otimes \left(\pi_0 \oplus \pi_1\right) \otimes \pi_{\text{NS}}$ coincides with the generating function for the number of pairs of Young diagrams with both white or both black corners.

The character of $\pi_0 \oplus \pi_1$ in principal grading is known to be \cite{Goddard:1986ee}

\be
\chi_{\pi_0 \oplus \pi_1} (q,t) = \Tr_{\pi_0 \oplus \pi_1}q^{L_0-{J^z_0 \over 2}}t^{J^z_0} = \chi_{{\cal F},NS}(q)\chi_{B}(q)\sum_{n\in {\mathbb Z}} q^{n(n-1) \over 2}t^n
\ee
One can derive this using formulae (\ref{rep relation NS}) and $L_0 = {a_0^2 \over 2 }+\sum_{n>0}a_{-n}a_n + \sum_{r>0} r \psi_{-r}\psi_r  $.

Then the character $\chi_{NS}(q,t)$ of the representation $\pi_H \otimes  (\pi_0 \oplus \pi_1) \otimes\pi_{NS}$ of the algebra ${\cal A}(2,2)$ in NS sector is just a product of the characters of each factor of this representation. So we obtain

\be
\label{A(2,2) character}
\chi_{NS}(q,t) = \chi^2_{{\cal F},NS}(q) \chi^3_B(q) \sum_{n\in {\mathbb Z}} q^{n(n-1) \over 2}t^n 
\ee
which is exactly the same as the generating function (\ref{diagram char NS}) for the number of diagrams with charges $(\sigma_1,\sigma_2) = (0,0)$ or $(1,1)$ and white and black cells.

\subsubsection{Character of ${\cal A}(2,2)$ representation in R sector}

In the R-sector we want to calculate the character 
$\chi_R(q,t) = \chi_{{\cal F},R}(q)\chi^2_{B}(q)\chi_{\pi_{1\over 2}}(q,t)$ 
and compare it with the generating function for the number of pairs of diagrams with charges 
$(\sigma_1 , \sigma_2) = (1,0)$ or $(0,1)$ with white and black cells.

The character $\chi_{\pi_{1\over 2}}(q,t)$ of the representation $\pi_{1\over 2}$ in principle grading is \cite{Goddard:1986ee}

\be
\chi_{\pi_{1\over 2}}(q,t) =  \Tr_{\pi_{1\over 2}}q^{L_0-{J^z_0 \over 2}}t^{J^z_0} = 
\chi_{{\cal F},R}(q)\chi_{B}(q)q^{1\over 16}\sum_{n \in {\mathbb Z}+{1\over 2}} q^{n(n-1) \over 2}t^{n}
\ee
One can evaluate this character using formulae (\ref{rep relation R}) and 
$L_0 = {1\over 16} + {a_0^2 \over 2 }+\sum_{n>0}a_{-n}a_n + \sum_{r>0} r \psi_{-r}\psi_r $.

The character of the representation $\pi_H \otimes   \pi_{1\over 2}\otimes\pi_{R}$ 
of the algebra ${\cal A}(2,2)$ in R-sector is the product of characters of each factor
\be
\chi_R(q,t) = \chi^2_{{\cal F},R}(q)\chi^3_{B}(q)q^{1\over 16}\sum_{n\in {\mathbb Z}+{1\over 2}} q^{n(n-1) \over 2}t^{n}
\ee
and it coincides with (\ref{diagram char R}) up to some unimportant factor.

The coincidence of the characters of the  algebra ${\cal A}(2,2)$ in NS- and R-sectors with the generating 
functions for the number of pairs of coloured diagrams with corresponding charges confirms our conjecture
about the AGT correspondence in the considered case. This fact suggests that there
exists a special basis in the conformal field theory which is enumerated in 
NS-sector by the pairs of Young diagrams with charges $(0,0)$ or $(1,1)$ and in R-sector with charges $(0,1)$ or $(1,0)$.

From this confirmation it follows that the fields from $ \hat sl(2)_2 $ can take part in
the relation between the instanton partition functions and conformal blocks which we are looking for.
Therefore in the next section we will write down the needed conformal blocks of $\hat{sl}(2)_2$ WZW fields.

\subsection{2-point and 4-point conformal blocks in $\hat{sl}(2)_2$ WZW model}

Let us denote by $\phi_{j}$ WZW primary fields corresponding to highest weight states $|j\r$ of the representation $\pi_j$.
It was shown in \cite{Belavin:2011sw} that to write an AGT representation for conformal blocks of 4 NS-fields one should take the composite field

\be
\label{NS vertex}
V_{NS} = \phi_{0} \otimes \Phi_{NS}
\ee
where $\phi_{0}={\mathbb 1}$ is WZW primary field with spin $j=0$. We extend this statement to R-sector as 

\be
\label{Ramond vertex}
V_{R} = \phi_{{1\over 2}} \otimes R
\ee
where  $\phi_{{1\over 2}}$ is WZW primary field with spin $j={1\over 2}$. So let us present some known results for
the 4-point function $\l \phi_{{1\over 2}}\phi_{{1\over 2}}\phi_{{1\over 2}}\phi_{{1\over 2}} \r$. 
Using the construction (\ref{free fields rep for currents1}), (\ref{free fields rep for currents2}) one could write

\be
\phi_{1\over 2} = \sigma e^{\phi}
\ee
where $\sigma$~-- the spin operator of Ising model. 
Conformal blocks corresponding to the correlation function $\l \phi_{{1\over 2}}\phi_{{1\over 2}}\phi_{{1\over 2}}\phi_{{1\over 2}} \r$
now are simply evaluated as the product of well-known conformal blocks of four spin operators and four exponents of scalar field. 

Since the dimension of $\phi_{1\over 2}$ is $\Delta_{1\over 2} = {3\over 16}$ (see (\ref{sl(2) weights})) the 2-point function of this field is

\be
\label{2-point WZW}
\l \phi_{1\over 2}(1) \phi_{1\over 2}(q) \r = {1\over (1-q)^{2\Delta}} = (1-q)^{-{3\over 8}} 
\ee

There are two 4-point conformal blocks of spin operators $\sigma$ \cite{Belavin:1984vu} so we have two 
corresponding $\hat{sl}(2)_2$ conformal blocks of fields $\phi_{1\over 2}$ (for the intermediate field with $\Delta = 0$ and $\Delta=1$ respectively)

\begin{align}
\label{WZW conf bl exp 1}
\l  \phi_{{1\over 2}}(\infty)\phi_{{1\over 2}}(1)\phi_{{1\over 2}}(q)\phi_{{1\over 2}}(0) \r_0  &= \left[ q(1-q) \right]^{-{3\over 8}}  \sqrt{1+\sqrt{1-q} \over 2} \simeq q^{-{3\over 8}} \left( 1+\frac{q}{4}+\frac{11 q^2}{64}+O[q]^3 \right) \\
\l  \phi_{{1\over 2}}(\infty)\phi_{{1\over 2}}(1)\phi_{{1\over 2}}(q)\phi_{{1\over 2}}(0) \r_1  &= \left[ q(1-q) \right]^{-{3\over 8}}\sqrt{1-\sqrt{1-q} \over 2} \simeq q^{{1\over 8}} \left( \frac{1}{2}+\frac{q}{4}+\frac{23 q^2}{128}+O[q]^{3} \right)
\label{WZW conf bl exp 2}
\end{align}

%%%%%%%%%%%%%%%%%%%%%%%%%%%%%%%%%%%%%%%%%%%%%%%%%%%%%%%%%%%%%%%%%%%%%%%%%%%%%%%%%%%%%%%%%%
%%%%%%%%%%%%%%%%%%%%%%%%%%%%%%%%%%%%%%%%%%%%%%%%%%%%%%%%%%%%%%%%%%%%%%%%%%%%%%%%%%%%%%%%%%
%%%%%%%%%%%%%%%%%%%%%%%%%%%%%%%%%%%%%%%%%%%%%%%%%%%%%%%%%%%%%%%%%%%%%%%%%%%%%%%%%%%%%%%%%%
%%%%%%%%%%%%%%%%%%%%%%%%%%%%%%%%%%%%%%%%%%%%%%%%%%%%%%%%%%%%%%%%%%%%%%%%%%%%%%%%%%%%%%%%%%

\section{SCFT and superconformal blocks}

\subsection{Neveu-Schwarz-Ramond algebra}

Let us consider $N=1$ superconformal field theory (SCFT). There are two conserved holomorphic currents 
in this theory: $T(z)$~-- energy-momentum tensor and $G(z)$~-- superconformal current. They are generators 
of conformal and superconformal transformations respectively. There are two types of fields in 
SCFT: Ramond (R) and Neveu-Schwarz (NS). We refer to them as R-fields and NS-fields. NS-fields are 
local and R-fields are half-local with respect to the supercurrent 
$G(z)$.

To construct the Hilbert space of SCFT one introduces Laurent modes which satisfy NSR algebra
 
\begin{align}
L_n &= \oint dz z^{n+1}T(z) \\
G_k &= \oint dz z^{k+1/2}G(z)
\end{align}
where due to the monodromy properties operators $L_n$ always appear with integer indices, and operators $G_k$ have integer indices in R-sector and half integer indices in NS-sector.

We use the following parametrizations of NS and R-fields dimensions and central charge

\begin{align}
\Delta_{NS} = {Q^2 \over 8}-{\lambda_{NS}^2 \over 2}, \qquad
\Delta_R &= {1\over 16} + {Q^2 \over 8}-{\lambda_{R}^2 \over 2}, \qquad \beta_{R} = {\lambda_{R} \over \sqrt{2}},\quad  \beta_{R} = {\lambda_{R} \over \sqrt{2}}, \\
c &= 1+2Q^2, \qquad Q = b+b^{-1}
\end{align}

Since there is the operator $G_0$ in R sector there must be two primary fields in this sector and they are related by the action of the operator $G_0$. We denote them $R^{\pm}$ and take the normalization as

\be \l R^{\e_1}(z_1)R^{\e_2}(z_2) \r = {\delta_{\e_1\e_2} \over (z_1 - z_2)^{2\Delta_R}}  \ee 

This dictates the choice 

\be G_0 R^{\pm} = i \beta_R e^{\mp i {\pi \over 4}}R^{\mp} \ee  
and also constraints parameter $\beta_R$ to be purely imaginary. We will take this fact into account when we will be evaluating conformal blocks. 

In what follows we will omit indices R and NS for parameters of critical dimensions $\lambda_{R},\lambda_{NS},\beta_{R},\beta_{NS}$ and for critical dimensions themselves.
Also we will need OPE's of holomorphic currents with R and NS fields

\begin{align}
\label{GF OPE}
G(z)\Phi(0) &\sim {\Psi(0) \over z} \\
G(z)\Psi(0) &\sim {2\Delta_{NS} \Phi(0) \over z^2} + {\p \Phi(0) \over z} \\
G(z)R^{\pm}(0) &\sim {i\beta e^{\mp i\pi\over 4} R^{\mp}(0) \over z^{3/2}} + {G_{-1}R^{\pm}(0) \over z^{1\over 2}} \label{GR OPE}
\end{align}
where $\Phi(u)$ is NS primary field with dimension $\Delta_{NS}$ and $\Psi(u)=G_{-1/2}\Phi(u)$. And also they have standard OPE's with energy-momentum tensor $T(z)$ as all primary fields have.
These OPEs are invariant under exchanging Ramond fields

\be
R^{\pm} \to e^{\mp i {\pi \over 4}} R^{\mp}
\ee

In the following subsections we use superconformal symmetry to derive the equations which define OPEs 
and conformal blocks with Ramond fields. 

Constraints on OPEs and conformal blocks of NS-fields were obtained some time ago in \cite{Belavin:2007zz}, \cite{Hadasz:2006qb} and on OPEs and conformal blocks including R-fields were obtained in \cite{Hadasz:2008dt}, \cite{Suchanek:2010kq}. However for completeness and convenience we rederive these constraints below in a slightly different way.      

%%%%%%%%%%%%%%%%%%%%%%%%%%%%%%%%%%%%%%%%%%%%%%%%%%%%%%%%%%%%%%%%%%%%%%%%%%%%%%%%%%%%%%%%%%
%%%%%%%%%%%%%%%%%%%%%%%%%%%%%%%%%%%%%%%%%%%%%%%%%%%%%%%%%%%%%%%%%%%%%%%%%%%%%%%%%%%%%%%%%%
%%%%%%%%%%%%%%%%%%%%%%%%%%%%%%%%%%%%%%%%%%%%%%%%%%%%%%%%%%%%%%%%%%%%%%%%%%%%%%%%%%%%%%%%%%
%%%%%%%%%%%%%%%%%%%%%%%%%%%%%%%%%%%%%%%%%%%%%%%%%%%%%%%%%%%%%%%%%%%%%%%%%%%%%%%%%%%%%%%%%%

\subsection{Superconformal blocks with two R-fields}

In this section our aim is to calculate conformal blocks corresponding to 4-point functions with two R fields and two NS fields. We are going to calculate them utilizing the method of "chain equations". 

Now it is time to decide what kind of conformal blocks we want to evaluate and we stress that we need the one corresponding to the 4-point function

\be 
\l \Phi(\infty)R(1)R(z)\Phi(0) \r
\ee

The bad thing about this correlator is that when one tries to evaluate its conformal blocks one needs to write chain equations for the OPE $R(z)\Phi(0)$. But in doing so one has to evaluate integrals of the type

\be
\oint du u^{k+{1\over 2}}G(u)R(z)\Phi(0)
\ee

In this integral a fractional singularity at $u=z$ appears due to (\ref{GR OPE}). Although there is a well known way to deal with it, here we want to follow much simpler way. In fact we will be dealing with the same exact problem a lot in further sections when correlators of four R-fields come up.

Nevertheless here it turns out to be much easier just to evaluate the conformal block of the 4-point function 
$\l R_2^{\pm}(\infty)\Phi_1(1)\Phi_4(z)R_3^{\pm}(0) \r$ first. Here no fractional powers appear in the integrals. After that we notice that of course we could relate this conformal block with the conformal block of $\l \Phi_1(\infty)R_2^{\pm}(1)R_3^{\pm}(z)\Phi_4(0) \r$ by some projective transformation. The projective transformation we need could be easily found from the fact that it should interchange points $1$ and $\infty$, and map point $z$ to $0$. The transformation we need is

\be
f(x)={x-z \over (1+{1-z \over L})x-1}
\ee
where $L$ ~-- is a regularization at infinity.
Under this transformation point $0$ maps to $z$ and thus using the standard transformation properties of primary fields we have the relation

\be \label{proj trans relation}
\l \Phi_1(\infty)R^{\pm}_2(1)R^{\pm}_3(z)\Phi_4(0) \r = (1-z)^{\Delta_1-\Delta_2+\Delta_3-\Delta_4}
\l  R^{\pm}_2(\infty)\Phi_1(1)\Phi_4(z)R^{\pm}_3(0)  \r
\ee

To work out the conformal block corresponding to 4-point function $\l  R^{\pm}_2(\infty)\Phi_1(1)\Phi_4(z)R^{\pm}_3(0)  \r$ we need the OPEs

\be
\label{chaindef1}
\Phi_2(z)R_1^{\pm}(0)=z^{\Delta-\Delta_1-\Delta_2} \sum_{N=0}^{\infty} z^N |N;1,2 \r_{\pm}  \\
\Psi_2(z)R_1^{\pm}(0) = z^{\Delta-\Delta_1-\Delta_2-{1\over 2}} \sum_{N=0}^{\infty} z^N \widetilde{|N;1,2 \r}_{\pm}
\label{chaindef2}
\ee

These equations account for contribution of one irreducible representation of NSR algebra with highest weight state $|\Delta\r$. In the case of OPE 
$\Phi_2(z)R_1^{\pm}(0)$ the highest weight state is $|\Delta\r = R^{\pm}(0)$ and in the case of OPE $\Psi_2(z)R_1^{\pm}(0)$ the highest weght state is
$|\Delta\r = R^{\mp}$. On the level $N$ there are contributions to $|N;1,2 \r_{\pm}$ and $\widetilde{|N;1,2 \r}_{\pm}$ from vectors $L_{-n_1}\dots L_{-n_s}G_{-r_1}\dots G_{-r_l}|\Delta\r$ with $\sum n_i + \sum r_i = N$.

States $|N;1,2 \r_{\pm}$ and $\widetilde{|N;1,2 \r}_{\pm}$ are usually referred to as chain vectors. In what follows we omit arguments $1,2$ and call them simply $|N \r_{\pm}$ and $\widetilde{|N \r}_{\pm}$ remembering that they depend on external fields dimensions and when computing conformal blocks we should remember that the bra state depend on $\Delta_1,\Delta_2$ and ket state depend on $\Delta_3,\Delta_4$.

Then we define the conformal blocks of the 4-point functions $\l R_2^{\pm}(\infty)\Phi_1(1)\Phi_4(z)R_3^{\pm}(0) \r$ as

\be
F^{\pm}(z)=\sum_{N=0}^{\infty} \prescript{}{\pm}\l N;1,2|N;3,4 \r_{\pm} z^N
\ee

Due to (\ref{proj trans relation}) the conformal block ${\cal F}^{\pm}(z)$ of the 4-point function $\l \Phi_1(\infty)R_2^{\pm}(1)R_3^{\pm}(z)\Phi_4(0) \r$ could be written as

\be
{\cal F}^{\pm}(z)=(1-z)^{\Delta_1-\Delta_2+\Delta_3-\Delta_4}F^{\pm}(z)
\ee

Now we proceed to deriving equations for $|N \r_{\pm}$.

Acting with $G_0$ on both sides of equations (\ref{chaindef1}) and using the OPEs (\ref{GF OPE}),(\ref{GR OPE}) we get

\begin{align}
& G_0 \left( \Phi_1(z) R_2^{\pm}(0) \right) = \oint du u^{k+{1\over 2}} G(u) \Phi_1(z) R_2^{\pm}(0)= \\
& \oint du u^{{1\over 2}} \left( {\Psi_1(z)R_2^{\pm}(0) \over u-z} + i\beta_2 e^{\mp i{\pi \over 4}} {\Phi_1(z)R_2^{\mp}(0) \over u^{3\over 2}} \right) = z^{1\over 2}\Psi(z)R_2^{\pm}(0) +i\beta_2 e^{\mp i{\pi \over 4}} \Phi_1(z)R_2^{\mp}(0)
\end{align}
No fractional singularities appear in the integrals here.
Substituting here OPE's from (\ref{chaindef1}), (\ref{chaindef2}) we get the constraint on these chain vectors

\be
\label{G0 constraint1}
G_0 |N \r_{\pm}=\widetilde{|N\r}_{\pm}+\beta_2 e^{\mp i \pi\over 4}|N\r_{\mp}\\
\ee

One may use this equation to express $\widetilde{|N\r}_{\pm}$ in terms of $|N \r_{\pm}$.

Acting with $G_0$ on (\ref{chaindef2}) gives the same constraint. 

Note that we has already implicitly used constraint from acting with $L_0$ to define $z$-dependence of the OPEs (\ref{chaindef1}), (\ref{chaindef2}).

Similarly acting with $G_k$ and $L_k$ ($k>0$) on the equations (\ref{chaindef1}),(\ref{chaindef2}) we obtain the constraints

\begin{align} 
\label{chain eq1}
			 G_k| N \r_{\pm} &= \widetilde{| N-k \r}_{\pm} \\
G_k \widetilde{| N \r}_{\pm} &= (2k\Delta_1-\Delta_2+\Delta+N-k)| N-k \r_{\pm} \\
			  L_k |N\r_{\pm} &= (k\Delta_1-\Delta_2+\Delta + N-k)|N-k\r_{\pm} \\
  L_k \widetilde{|N\r}_{\pm} &= (k(\Delta_1+{1\over 2})-\Delta_2+\Delta+N-k)\widetilde{|N-k\r}_{\pm} \label{chain eq2}
\end{align}

Actually commutation relations (\ref{NSR algebra LL}) - (\ref{NSR algebra LG}) imply that we only need these equations for $G_1$ and $L_1$ since $G_1$ and $L_1$ generate all the other operators $G_k$ and $L_k$ for $k>0$ due to commutation relations.

To evaluate $|N\r_{\pm}$ we get rid of $\widetilde{|N\r}_{\pm}$ with the help of the equation (\ref{G0 constraint1})

\be
\widetilde{|N\r}_{\pm}=G_0 |N \r_{\pm}-i\beta_2 e^{\mp i \pi\over 4}|N\r_{\mp}
\ee

From this relation and equations (\ref{chain eq1})-(\ref{chain eq2}) we get two independent equations for $|N\r_{\pm}$

\begin{align}
\label{main eq1}
& L_1 |N \r_{\pm}=(\Delta+\Delta_1-\Delta_2+\Delta+N-1)|N-1\r_{\pm} \\
& G_1 |N \r_{\pm}=G_0|N-1\r_{\pm}-i\beta_2 e^{\mp i {\pi\over 4}}|N-1\r_{\mp}
\label{main eq2}
\end{align}

As was mentioned it is enough to take only constraints with $L_1$ and $G_1$.

These equations allow us evaluate chain vectors $|N\r_{\pm}$ and conformal blocks corresponding to the correlators $\l R_1^{\pm}(\infty) \Phi_2(1) \Phi_3(z) R_4^{\pm}(0) \r$ and consequently conformal blocks for the correlators $\l \Phi_2(\infty)R_1^{\pm}(1)R_4^{\pm}(z)\Phi_3(0) \r$.
Some explicit formulae for chain vectors on low levels could be found in appendix A.

\subsection{Superconformal blocks of four R-fields}

\subsubsection{OPE of two Ramond fields}

In this section we switch to considering the 4-point functions of Ramond fields $\l RRRR \r$. 
The OPEs for two Ramond fields could be written as follows

\be
R_1^{\pm}(z)R_2^{\pm}(0)= z^{\Delta-\Delta_1 - \Delta_2} \sum_{N=0}^{\infty} z^N | N ; \pm\pm \r \\
R_1^{\pm}(z)R_2^{\mp}(0)= z^{\Delta-\Delta_1 - \Delta_2} \sum_{N=0}^{\infty} z^N | N ; \pm\mp \r
\ee

It is not possible here to derive ordinary chain equations in the same way as we did in the case of 4-point function $\l R(\infty)\Phi(1)\Phi(z)R(0) \r$. The situation here is similar to that of $\l \Phi(\infty)R(1)R(z)\Phi(0) \r$. When acting by generators $G_r$ on the OPE of two R-fields one cannot deform the contour around both R-fields to contours around each R-field and act with the generator $G_r$ on each R-field separately because of the cut between two R-fields. Nevertheless it is possible to derive constraints on chain vectors in this case too.
Let us consider the contour integrals around points $z,0$

\be
\oint du \sqrt{u(u-z)}u^{k+{1\over 2}}G(u)R_1^{\pm}(z)R_2^{\pm}(0) \\
\oint du \sqrt{u(u-z)}u^{k+{1\over 2}}G(u)R_1^{\pm}(z)R_2^{\mp}(0)
\ee   

Evaluating these integrals in two ways either by expanding the square root around $u=0$ or using the OPEs of supercurrent with R-fields (\ref{GR OPE}) we get the chain equations

\be
\sum_{m=0}^{[N]} (-1)^m C_{1\over 2}^m G_{k-m+1}|N-m;\pm\pm \r = i\beta_1 e^{\mp i{\pi \over 4}} |N-k-1;\mp\pm \r  \mp 
\delta_{k,-{1\over 2}} \beta_2 e^{\mp i{\pi \over 4}} |N-{1\over 2};\pm\mp \r
\label{chain1 RRRR}  \\
\sum_{m=0}^{[N]} (-1)^m C_{1\over 2}^m G_{k-m+1}|N-m;\pm\mp \r = i\beta_1 e^{\mp i{\pi \over 4}} |N-k-1;\mp\mp \r  \mp
\delta_{k,-{1\over 2}} \beta_2 e^{\pm i{\pi \over 4}} |N-{1\over 2};\pm\pm \r
\label{chain2 RRRR}
\ee
where $[N]$ is an integer part of $N$ and $C_{1\over 2}^m$ denotes the binomial coefficient.

	Notice that states $| 0; \pm\mp \r,|{1\over 2}: \pm\pm \r $ vanish since they correspond to three point functions $\l \Phi R^{\pm} R^{\mp} \r$ and $\l \Psi R^{\pm} R^{\pm} \r$ which have fermionic number one. Similarly all states $|n;\pm\mp \r$ vanish for $n$~-- integer and $|r;\pm\pm \r$ vanish for $r$~-- half integer.

On the zeroth level there is of course only one state, namely NS-state with dimension $\Delta$. Thus states  $| 0; \pm\pm \r$ should be proportional to this NS-state.

\be
| 0; \pm\pm \r = \gamma_{\pm} |0\r
\ee

\subsubsection{Diagonalizing chain equations}
	One notices that equations (\ref{chain1 RRRR})-(\ref{chain2 RRRR}) do not totally define vectors $| 0; \pm\pm \r$ since they include two independent constants $\gamma_{\pm}$ in a non-trivial way. Thus we will "diagonalize" these equations in the sense that we will define some linear combinations of $| N; \pm\pm \r$ and $| N; \pm\mp \r$ in a way that they simply depend on $\gamma_{\pm}$. Namely they will include $\gamma_{\pm}$ as an overall factor.  

Taking the chain equations for $k=-{1\over 2}$ we get 

\begin{align}
& |N- {1\over 2} ; \pm \mp \r = \pm {e^{\pm i {\pi \over 4}} \over \Delta_2 - \Delta_1} \sum_{m=0}^{[N]}(-)^m C_{1\over 2}^m G_{{1\over 2}-m} 
\Big( \beta_2|N-m ; \pm\pm \r - \beta_1 |N-m;\mp\mp \r \Big) 
\label{1k=1/2}
\\
& |N- {1\over 2} ; \pm\pm \r = \pm {e^{ \mp i {\pi \over 4}} \over \Delta_2 - \Delta_1} \sum_{m=0}^{[N]}(-)^m C_{1\over 2}^m G_{{1\over 2}-m} 
\Big(\beta_2|N-m ; \pm \mp \r \mp i\beta_1 |N-m;\mp\pm\r \Big)
\label{2k=1/2}
\end{align}

Plugging this into chain equations for $k \geq {1\over 2}$ we get

\begin{align}
& \sum_{m=0}^{[N]}(-)^m C_{1\over 2}^m G_{k-m+1} |N-m;\pm \pm\r = \nn \\ &= {\beta_1 \over \Delta_2-\Delta_1}\sum_{m=0}^{[N]-k-{1\over 2}}(-)^m C_{1\over 2}^m G_{{1\over 2}-m}
\left( \beta_1|N-m-k-{1\over 2} ; \pm \pm \r - \beta_2 |N-m-k-{1\over 2};\mp\mp\r \right)
\label{prediag 1}
\\
& \sum_{m=0}^{[N]}(-)^m C_{1\over 2}^m G_{k-m+1} |N-m;\pm \mp\r = \nn \\ &= {\beta_1 \over \Delta_2-\Delta_1}\sum_{m=0}^{[N]-k-{1\over 2}}(-)^m C_{1\over 2}^m G_{{1\over 2}-m}
\left( \beta_1|N-m-k-{1\over 2} ; \pm \mp \r \mp i\beta_2 |N-m-k-{1\over 2};\mp\pm\r \right)
\label{prediag 2}
\end{align}

To diagonalize these equations let us define

\begin{align}
|N;\pm\r_ = \begin{cases}  |N;++\r \pm |N;--\r  \quad \text{if $N \in {\mathbb Z}$} \\
 							|N;+-\r \mp i|N;-+\r \quad \text{if $N \in {\mathbb Z}+{1\over 2}$}
\end{cases}
\end{align}

Taking the pair of equations (\ref{prediag 1}) and adding them to each other or subtracting from each other and doing the same with the pair of equations (\ref{prediag 2}) we are able to diagonalize these equations $(k \geq {1\over 2})$

\be
\label{RRRR main1}
\sum_{m=0}^{[N]}(-)^m C_{1\over 2}^m G_{k-m+1} |N-m;\pm\r = {\beta_1 \over \beta_1 \pm (-1)^{2N}\beta_2}
\sum_{m=0}^{[N]-k-{1\over 2}}(-)^m C_{1\over 2}^m G_{{1\over 2}-m} |N-m-k-{1\over 2} ;\pm\r
\ee
where $N \in {1\over 2}{\mathbb Z}$.
Also here we made use of the fact that $\Delta_2 - \Delta_1 = \beta_1^2 - \beta_2^2$.

To have the full system of equations for states $|N;\pm\r$ one can use equations (\ref{1k=1/2}) and (\ref{2k=1/2}) to get

\begin{align}
\label{RRRR main2}
|N;\pm\r =  \pm (-1)^{2N} {e^{ - i {\pi \over 4}(-1)^{2N}} \over \beta_1 \mp (-1)^{2N} \beta_2} \sum_{m=0}^{[N]}(-)^m C_{1\over 2}^m G_{{1\over 2}-m} 
|N-m+{1\over 2};\pm\r
\end{align}

Equations (\ref{RRRR main1}), (\ref{RRRR main2}) form the full system which allows to determine states $|N\r_{\pm}$ on all levels.
These equations are diagonalized in the sense that the solution could be found in the form

\begin{align}
|N;\pm\r &= Y_{\pm}(N)|0;\pm\r 
\end{align}
where $Y_{\pm}(N)$ is some operator on the level N constructed out of operators $L_n, G_k$. So as was promised vectors $|N;\pm\r $ include constants $\gamma_{\pm}$ only in combinations $\Gamma_{\pm}=(\gamma_{+}\pm\gamma_{-})$ and only as an overall factor since $|0;\pm\r = \Gamma_{\pm} |0\r$.

%%%%%%%%%%%%%%%%%%%%%%%%%%%%%%%%%%%%%%%%%%%%%%%%%%%%
%%%%%%%%%%%%%%%%%%%%%%%%%%%%%%%%%%%%%%%%%%%%%%%%%%%%
%%%%%%%%%%%%%%%%%%%%%%%%%%%%%%%%%%%%%%%%%%%%%%%%%%%%
%%%%%%%%%%%%%%%%%%%%%%%%%%%%%%%%%%%%%%%%%%%%%%%%%%%%

\section{Comparison of superconformal blocks with partition functions on ${\mathbb R}^4/{\mathbb Z}_2$}

First of all we review the instanton counting on ${\mathbb R}^4/{\mathbb Z}_2$. Let us consider $N=2$ SU(2) gauge theory on ${\mathbb R}^4/{\mathbb Z}_2$.  
The partition function for four fundamental hypermultiplets for this theory is 

\be
Z^{(q_1,q_2)}_{(u_1,u_2),(v_1,v_2)}(P_{1},\alpha_{1},\alpha_{2},P_{2}|P|z)=
\sum_{\{\vec{Y}^{\vec{q}}\}}
\frac{Z_{\text{bif}}(\alpha_{1}|\vec{P}_{1},\vec{\varnothing}^{u},\vec{P},
\vec{Y^{q}})
Z_{\text{bif}}(\alpha_{2}|\vec{P},\vec{Y^{q}},\vec{P}_{2},
\vec{\varnothing}^{v})}
{Z_{\text{bif}}(0|\vec{P},\vec{Y^{q}},\vec{P},
\vec{Y^{q}})}\cdot z^{\frac{|\vec{Y^{q}}|}{2}} \label{partition function}
\ee
where the sum goes over the pairs of chess coloured Young diagrams $\vec{Y} = (Y_1^{q_1},Y_2^{q_2})$. 
One ascribes a ${\mathbb Z}_{2}$ charge $q_{i}$, $i =1,2$ to each diagram which is just the color of the corner and it takes values 
0 for white, 1 for black cells. Vector $\vec{P} = (P,-P)$ stands for the vev of the scalar field. The parameters $\alpha_1, \alpha_2, P_1, P_2$ are related to masses of fundamental hypermultiplets \cite{Alfimov:2011ju}. The symbol $|\vec{Y^{q}}|$ stands for the total number of cells in both diagrams.

The bifundamental multiplet contribution to the partition function of $SU(2)$ gauge theory on
${\mathbb R}_4/{\mathbb Z}_2$ is defined as
\begin{align}
Z_{\text{bif}}(\alpha|\vec{P}',\vec{W^{u}},\vec{P},\vec{Y^{q}})
= \prod_{i,j=1}^{2} \prod_{s\in Y_{i}^{q_{i}\heartsuit}}
(Q-E_{Y_{i},W_{j}}(P_{i}-P'_{j}|s)-\alpha)\prod_{t\in
W_{j}^{u_{j}\heartsuit}}(E_{W_{j},Y_{i}}(P'_{j}-P_{i}|t)-\alpha),
\label{Zbif}
\end{align}
where
$$
E_{Y,W}(P|s)= P- l_{W}(s)b^{-1} +(a_{Y}(s)+1)b
$$
and the product over  $s \in Y_{i}^{q_{i}}$
and $t \in W_{j}^{u_{j}}$ is limited to the set $\heartsuit$ that
includes all $s, t$ satisfying the equations
\begin{align}
&s \in  Y_{i}^{q_{i}\heartsuit}:\;l_{W_{j}}(s)+a_{Y_{i}}(s)+1 \equiv u_{j}-q_{i} \; \textrm{mod} \; 2, \notag \\
&t \in  W_{j}^{u_{j}\heartsuit}:\;l_{Y_{i}}(t)+a_{W_{j}}(t)+1 \equiv q_{i}-u_{j} \; \textrm{mod} \; 2,
\end{align}
Notice that these conditions are invariant under flipping of all charges.

In order to identify partition functions with conformal blocks one needs to take the sum in (\ref{partition function}) over special class of diagrams $\{{\vec Y}^{q}\}$ depending on the type of the intermediate state (NS or R state) in the considered conformal block. Namely for NS-sector we have to take either diagrams with equal number of black and white cells and with charges $(0,0)$ or diagrams with the difference between black and white cells equal to one and charges $(1,1)$ \cite{Belavin:2011pp, Belavin:2011tb}. And in R-sector we take diagrams with equal number of black and white cells and charges $(0,1)$ or $(1,0)$ \cite{Ito:2011mw}.

Thus here we assume the same relation between the charges of the diagrams and representations of the algebra ${\cal A}(2,2)$. Let us however emphasize that  identification of conformal blocks with partition functions requires additional restriction on the number of white and black cells in the Young diagrams.

To compare partition functions with conformal blocks we as usually make identifications

\be
P_1=\lambda_1; \quad \alpha_1 = {Q\over 2}+\lambda_2; \quad \alpha_2 = {Q\over 2} + \lambda_3; \quad P_2 = \lambda_4; \quad P = \lambda
\ee

Also we omit most of the variables which the partition function depend on and use a shorter notation $\prescript{\diamondsuit}{}{Z}^{(q_1,q_2)}_{(u_1,u_2)(v_1,v_2)}$ or $\prescript{\vardiamond}{}{Z}^{(q_1,q_2)}_{(u_1,u_2)(v_1,v_2)}$ where $(q_1,q_2)$~-- charges of both pair of non-empty diagrams, $(u_1,u_2)$~-- charges of the first pair of empty diagrams, $(v_1,v_2)$~-- charges of the second pair of empty diagrams and left upper indices $^{\diamondsuit}$ and $ ^{\vardiamond}$ stand for the difference between black and white cells to be $0$ and $1$ respectively.

Following \cite{Alday:2010vg}, \cite{Alba:2010qc} we will assume that the identification of the partition functions with the conformal blocks is performed in the following way. We suggest that there exists an orthogonal basis $|P\r_{\vec{Y}^{q}}$ in the representation space of the ${\cal A}(2,2)$ algebra with the scalar product

\be
\prescript{}{\vec{Y}^q}\l P_1|P_2  \r_{\vec{W}^u} = \delta_{\vec{Y}^q, \vec{W}^u} Z_{\textrm{bif}}(0|\vec{P}_1,\vec{Y^{q}},\vec{P}_2,
\vec{Y^{q}})
\ee

Matrix elements of the vertex operator in this basis is given by 
the contribution of the bifundamental multiplet

\be
\label{3-point func}
{\prescript{}{\vec{Y}^{q}}\l P_1| V_{\alpha} | P_2 \r_{\vec{W}^{u}}  } =
Z_{\textrm{bif}}(\alpha|\vec{P}_1,\vec{Y}^{q},\vec{P}_2,\vec{W}^{u})
\ee

Consider the 4-point function $\l V_{P_1}(\infty)V_{\alpha_1}(1)V_{\alpha_2}(z)V_{P_2}(0)\r$. 
The primary state corresponds to the basis vector which is labelled by the pair of empty diagrams. These diagrams have the charges $\vec{u}=(u_1,u_2)$ depending on the type of the primary state (NS or R). So one can write 
\be
\l V_{P_1}(\infty)V_{\alpha_1}(1)V_{\alpha_2}(z)V_{P_2}(0)\r = 
\prescript{}{\vec{\varnothing}^u}\l P_1|V_{\alpha_1}(1)V_{\alpha_2}(z) |P_2 \r_{\vec{\varnothing}^v} 
\ee

To evaluate this 4-point function one inserts a full set of orthogonal states 
\be
{\mathbb 1}=\sum_{\{\vec{Y}^q\}} |P\r_{\vec{Y}^{q}} {1\over \prescript{}{\vec{Y}^q}\l P|P  \r_{\vec{Y}^q} } \prescript{}{\vec{Y}^q} \l P|
\ee
in the middle of the 4-point function. One gets expression

\be
\label{part}
\prescript{}{\vec{\varnothing}^u}\l P_1|V_{\alpha_1}(1)V_{\alpha_2}(z) |P_2 \r_{\vec{\varnothing}^v} = 
\sum_{\{\vec{Y}^{\vec{q}}\}}
\frac{Z_{\textrm{bif}}(\alpha_{1}|\vec{P}_1,({\varnothing}^{u_1},{\varnothing}^{u_2}),\vec{P},
\vec{Y^{q}})
Z_{\textrm{bif}}(\alpha_{2}|\vec{P},\vec{Y^{q}},\vec{P}_2,
({\varnothing}^{v_1},{\varnothing}^{v_2}))}
{Z_{\textrm{bif}}(0|\vec{P},\vec{Y^{q}},\vec{P},
\vec{Y^{q}})}\cdot z^{\frac{|\vec{Y^{q}}|}{2}}
\ee

Notice that empty diagrams 
$\vec{\varnothing}^u, \vec{\varnothing}^v$ in $Z_{\textrm{bif}}$ correspond to fields 
at infinity and at zero ~-- $V_{P_1}(\infty)$ and $V_{P_2}(0)$.

In section 2 we have seen that the NS-sector character coincides with 
the generating function for pairs of Young diagrams with charges $(0,0)$ and $(1,1)$ 
and R-sector character coincides with generating function for the pairs
of diagrams with charges $(1,0)$ and $(0,1)$. From this one can anticipate what kind of
partition functions one should take in order to compare it with conformal blocks.

Firstly let us consider the 4-point function $\l \Phi(\infty) R(1) R(z) \Phi(0) \r$. 
The intermediate states here consist of Ramond fields. 
So we conclude that the sum on the right hand side of (\ref{part}) should be taken only over the
diagrams with charges $(1,0)$ or $(0,1)$. Also the states $\Phi(\infty)$ and $\Phi(0)$ are NS-fields 
and thus the empty diagrams $\vec{\varnothing}^u, \vec{\varnothing}^v$ in (\ref{part}) should have charges $(0,0)$ or $(1,1)$. 

In the case of 4-point function of four Ramond fields $\l R(\infty)R(1)R(z)R(0) \r$ 
the intermediate field is NS-field so the sum in the partition function should be taken 
only over diagrams with charges $(0,0)$ or $(1,1)$. The fields at infinity and at zero are 
$R(\infty)$ and $R(0)$~-- R-fields so the charges of the empty diagrams 
in $Z_{\textrm{bif}}$ in (\ref{part}) should be either $(1,0)$ or $(0,1)$.

\subsection{Superconformal blocks $\l \Phi R R \Phi \r$}

Taking into account the discussion in the previous subsection it is natural to propose that  

\begin{align}
\label{main conj R-sector}
\prescript{\diamondsuit}{}Z^{(0,1)}_{(0,0)(0,0)}(q) = (1-q)^U (1-q)^{-{3\over 8}}  {\cal F}^{+}(q) 
\end{align}
where $U =({Q\over 2} + \lambda_2)({Q\over 2} - \lambda_3)$ and the left upper index $^\diamondsuit$ denotes that the sum in the partition function is taken only over the diagrams with equal 
number of white and black cells.
We checked this relation on the first two levels. Some explicit formulae on these levels can be found in the appendix A.
Notice that the factor $(1-q)^{-{3\over 8}}$ is exactly the 2-point function (\ref{2-point WZW}) of  $\hat{sl}(2)_2$ WZW-field with spin $j={1\over 2}$ which has weight ${3\over 16}$. 

This confirms the proposition (\ref{Ramond vertex}) that Ramond vertex operator ${\mathbb V}_R$ includes the field 
$\phi_{1\over 2}$ from $\hat{sl}(2)_2$ WZW model

\be
\label{full Ramond vertex}
{\mathbb V}_R = {\cal V} \otimes \phi_{1\over 2} \otimes R
\ee
where ${\cal V}$ is Carlsson-Okounkov vertex operator \cite{Alba:2010qc, Carlsson} of Heisenberg factor ${\cal H}$ and $R$ is Ramond primary field of NSR factor of algebra $ {\cal H} \oplus \hat{sl}(2)_2 \oplus \text{NSR}$.

\subsection{Superconformal blocks $\l RRRR \r$}

In the case of four R-fields there are many different conformal blocks one could evaluate. There are vectors $|N;\pm\r$ on the integer and half-integer levels. So we define different conformal blocks as

\begin{align}
F_{\pm} = {1\over \Gamma_{\pm} \Gamma_{\pm}}\sum_{N=0,1,\dots} q^N \l N;\pm|N;\pm \r, \quad 
H_{\pm} ={1\over \Gamma_{\pm} \Gamma_{\mp}} \sum_{N=0,1,\dots} q^N \l N;\pm|N;\mp \r \\
\widetilde{F}_{\pm} = {(-i)\over \Gamma_{\pm} \Gamma_{\pm}}\sum_{N={1\over 2},{3\over 2},\dots} q^{N} \l N;\pm|N;\pm \r, \quad 
\widetilde{H}_{\pm} = {1\over \Gamma_{\pm} \Gamma_{\mp}}\sum_{N={1\over 2},{3\over 2},\dots} q^{N} \l N;\pm|N;\mp \r
\end{align}
where conformal blocks are divided by $\Gamma_{\pm}$ so if we take normalization $\l0|0\r = 1$ then conformal blocks on $F_{\pm}$ have the form $1+F_{\pm 1}q+\dots$.

We stress that for partition functions with equal numbers of white and black cells

\be
\label{main conj NS-sector 1}
\prescript{\diamondsuit}{}{Z}^{(0,0)}_{(1,0)(1,0)}(q) &= (1-q)^U \left( G_{sl(2)}(q) H_{-}(q) +  \widetilde{G}_{sl(2)}(q) \widetilde{H}_{-}(q) \right) \\
\prescript{\diamondsuit}{}{Z}^{(0,0)}_{(0,1)(0,1)}(q) &= (1-q)^U \left( G_{sl(2)}(q) H_{+}(q) +  \widetilde{G}_{sl(2)}(q) \widetilde{H}_{+}(q) \right) \\
\prescript{\diamondsuit}{}{Z}^{(0,0)}_{(1,0)(0,1)}(q) &= (1-q)^U \left( G_{sl(2)}(q) F_{-}(q) +  \widetilde{G}_{sl(2)}(q) \widetilde{F}_{-}(-q) \right) \\
\prescript{\diamondsuit}{}{Z}^{(0,0)}_{(0,1)(1,0)}(q) &= (1-q)^U \left( G_{sl(2)}(q) F_{+}(q) +  \widetilde{G}_{sl(2)}(q) \widetilde{F}_{+}(-q) \right)
\ee

And for partition functions with the difference between black and white cells equal to one

\begin{align}
\prescript{\vardiamond}{}Z^{(1,1)}_{(1,0)(1,0)}(q) &= (1-q)^U \left( \widetilde{G}_{sl(2)}(q) H_{+}(q) +  G_{sl(2)}(q) \widetilde{H}_{+}(q)   \right) \\
\prescript{\vardiamond}{}Z^{(1,1)}_{(0,1)(0,1)}(q) &= (1-q)^U \left( \widetilde{G}_{sl(2)}(q) H_{-}(q) +  G_{sl(2)}(q) \widetilde{H}_{-}(q)   \right) \\
\prescript{\vardiamond}{}Z^{(1,1)}_{(1,0)(0,1)}(q) &= (1-q)^U \left( \widetilde{G}_{sl(2)}(q) F_{+}(q) +  G_{sl(2)}(q) \widetilde{F}_{+}(-q) \right) \\
\prescript{\vardiamond}{}Z^{(1,1)}_{(0,1)(1,0)}(q) &= (1-q)^U \left( \widetilde{G}_{sl(2)}(q) F_{-}(q) +  G_{sl(2)}(q) \widetilde{F}_{-}(-q) \right)
\label{main conj NS-sector 2}
\end{align}
Again we checked these relations on the first two levels and some of these formulae can be found in the appendix B. From these comparisons we found first coefficients of $\hat{sl}(2)_2$ factors

\begin{align}
G_{sl(2)} &= 1+{1\over 4}q + {11\over 64} q^2 + \dots \\
\widetilde{G}_{sl(2)} &= {1\over 2} q^{1\over 2} + {1\over 4} q^{3\over 2} + \dots 
\end{align}
which is the same as in (\ref{WZW conf bl exp 1}), (\ref{WZW conf bl exp 2}).

Thus the formulae (\ref{main conj NS-sector 1})~--(\ref{main conj NS-sector 2}) again confirm the expression (\ref{full Ramond vertex}) for the Ramond vertex operator.

\begin{comment}

\subsection{SL(2) fields 4-point function}

\begin{align} 
8 A' &= {1\over x} A+{1\over 1-x}(3A+2B) \\
8 B' &= -{1\over x} (2A+3B)-{1\over 1-x} B
\end{align}

\begin{align}
A &= -4 x B' -B(1+{1\over 2(1-x)}) \\
B &= 4(1-x)A' -A(1+{1\over 2x})
\end{align}

Solutions are

\begin{align}
A_0(x) &= {1\over 2}x^{5\over 8} (1-x)^{-{3\over 8}} F\left( {3\over 4},{1\over 4},{3\over 2},x \right) = x^{-{3\over 8}} \left( \frac{x}{2}+\frac{x^2}{4}+\frac{23 x^3}{128}+O[x]^4 \right)  \\
B_0(x) &= x^{-{3\over 8}} (1-x)^{{1\over 8}} F\left( {1\over 4},-{1\over 4},{1\over 2},x \right) =x^{-{3\over 8}}\left( 1-\frac{x}{4}-\frac{5 x^2}{64}-\frac{11 x^3}{256}+O[x]^{4} \right) \\
A_1(x) &= x^{1\over 8} (1-x)^{-{3\over 8}} F\left( {1\over 4},-{1\over 4},{1\over 2},x \right) = x^{1\over 8}\left( 1+\frac{x}{4}+\frac{11 x^2}{64}+\frac{35 x^3}{256}+O[x]^{4} \right) \\
B_1(x) &= -{1\over 2}x^{1\over 8} (1-x)^{{1\over 8}} F\left( {3\over 4},{1\over 4},{3\over 2},x \right) = x^{1\over 8} \left( -\frac{1}{2}+\frac{x^2}{128}+\frac{x^3}{128}+O[x]^4 \right)
\end{align}

\end{comment}

\section{Concluding remarks}
In this paper we extended the AGT relation between $N=2$ SUSY Yang-Mills theory on ${\mathbb R}^4/{\mathbb Z}_2$ and CFT model with the algebra ${\cal A}(2,2)={\cal H}\oplus \hat{sl}(2)_2\oplus \text{NSR}$ to Ramond sector. The main results are the explicit expressions (\ref{main conj R-sector}), (\ref{main conj NS-sector 1})~--(\ref{main conj NS-sector 2}) of conformal blocks with Ramond fields in terms of Instanton partition functions. These formulae are the first examples of non-trivial manifestation of the $\hat{sl}(2)_2$ part of ${\cal A}(2,2)$ algebra.  

In \cite{Bonelli:2011kv}, \cite{Belavin:2011sw} it was shown that 3-point and 4-point correlation functions in super Liouville field theory have a representation in terms of the product of two ordinary Liouville field theories. In \cite{Schomerus:2012se} this was extended to 3-point functions with Ramond fields. It would be interesting to find a similar construction in the case of 4-point conformal blocks with Ramond fields considered here.

\section*{Acknowledgements}
The authors thank V.Belavin, M.Bershtein and G.Tarnopolsky for useful discussions. We also thank
S. Pugai for critical reading of the manuscript and useful comments. We
are grateful to G.Tarnopolsky for sharing with us some powerful Mathematica
files devoted to instanton computations.
This work was supported by RFBR grant No.12-01-00836-A, by the Russian Ministry of Education and Science under the grants 2012-1.5-12-000-1011-012, contract No.8528 and 2012-1.1-12-000-1011-016, contract No.8410 and  partially by RFBR research project No.12-02-33011. The work of B.M. was also supported by Dynasty foundation grant.

%%%%%%%%%%%%%%%%%%%%%%%%%%%%%%%%%%%%%%%%%%%%%%%%%%%%%%%%%%%%%%%%%%%%%%%%
%%%%%%%%%%%%%%%%%%%%%%%%%%%%%%%%%%%%%%%%%%%%%%%%%%%%%%%%%%%%%%%%%%%%%%%%
%%%%%%%%%%%%%%%%%%%%%%%%%%%%%%%%%%%%%%%%%%%%%%%%%%%%%%%%%%%%%%%%%%%%%%%%
\Appendix
%%%%%%%%%%%%%%%%%%%%%%%%%%%%%%%%%%%%%%%%%%%%%%%%%%%%%%%%%%%%%%%%%%%%%%%%
%%%%%%%%%%%%%%%%%%%%%%%%%%%%%%%%%%%%%%%%%%%%%%%%%%%%%%%%%%%%%%%%%%%%%%%%
%%%%%%%%%%%%%%%%%%%%%%%%%%%%%%%%%%%%%%%%%%%%%%%%%%%%%%%%%%%%%%%%%%%%%%%%

\section{Comparison of $\l \Phi R R \Phi \r$ superconformal block with partition functions on the first two levels}

\subsection{First level}

On the 0th and 1st levels we have chain vectors

\begin{align}
& \text{0th level:} \quad    |0 \r_{\pm} =|R^{\pm} \r  \\
& \text{1st level:} \quad    |1\r_{\pm} = z_1^{\pm} L_{-1}|0\r_{\pm} + z_2^{\pm} G_{-1}|0 \r_{\mp}
\end{align}

Solving equations (\ref{main eq1}),(\ref{main eq2}) we get the answer on the 1st level

\begin{align}
& z_1^{\pm} = \frac{12 \beta ^2-12 \beta  \text{$\beta $}_2+(3 c+16 \Delta ) (\Delta +\text{$\Delta $}_1-\text{$\Delta $}_2)}{18 \beta ^2+6 c \Delta +32 \Delta ^2}  \\
& z_2^{\pm} = \pm \frac{2 e^{\pm i {\pi\over 4}} (-4 \text{$\beta $}_2 \Delta +\beta  (\Delta -3 \text{$\Delta $}_1+3 \text{$\Delta $}_2))}{9 \beta ^2+\Delta  (3 c+16 \Delta )}
\end{align}

We found the conformal block on the first level

\begin{align}
F_1^{+} = F_1^{-}  = & \frac{(3 c+16 \Delta ) (\Delta -\text{$\Delta $}_1+\text{$\Delta $}_2) (\Delta +\text{$\Delta $}_3-\text{$\Delta $}_4)}{18 \beta ^2+6 c \Delta +32 \Delta ^2} +
\frac{4 \beta ^2 (2 \Delta -3 (\text{$\Delta $}_1-\text{$\Delta $}_2-\text{$\Delta $}_3+\text{$\Delta $}_4))}{18 \beta ^2+6 c \Delta +32 \Delta ^2} + \nn \\
& + \frac{4 \beta  \text{$\beta $}_3 (\Delta +3 \text{$\Delta $}_1-3 \text{$\Delta $}_2)+4 \beta \text{$\beta $}_2 (\Delta -3 \text{$\Delta $}_3+3 \text{$\Delta $}_4)-16 \text{$\beta $}_2 \text{$\beta $}_3 \Delta}{18 \beta ^2+6 c \Delta +32 \Delta ^2}
\end{align}

Our aim is to compare this norm with partition function with charge (0,1) on the first level 

\small
\begin{align} 
&\prescript{\diamondsuit}{1}Z^{(0,1)}_{(0,0),(0,0)}=\\ 
& \frac{\left(2 a+3 \epsilon _1+\epsilon _2-2 \lambda _1+2 \lambda _2\right) \left(2 a+3 \epsilon _1+\epsilon _2+2 \lambda _1+2 \lambda _2\right) \left(2 a+3 \epsilon _1+\epsilon _2-2 \lambda _3-2 \lambda _4\right) \left(2 a+3 \epsilon _1+\epsilon _2-2 \lambda _3+2 \lambda _4\right)}{32 \epsilon _1 \left(2 a+\epsilon _1\right) \left(\epsilon _1-\epsilon _2\right) \left(2 a+2 \epsilon _1+\epsilon _2\right)} \nn \\
%%%%%%%%%%%%%%%%%%%%%%%%%%
+ & \frac{\left(2 a+\epsilon _1+3 \epsilon _2-2 \lambda _1+2 \lambda _2\right) \left(2 a+\epsilon _1+3 \epsilon _2+2 \lambda _1+2 \lambda _2\right) \left(2 a+\epsilon _1+3 \epsilon _2-2 \lambda _3-2 \lambda _4\right) \left(2 a+\epsilon _1+3 \epsilon _2-2 \lambda _3+2 \lambda _4\right)}{32 \epsilon _2 \left(2 a+\epsilon _2\right) \left(-\epsilon _1+\epsilon _2\right) \left(2 a+\epsilon _1+2 \epsilon _2\right)}  \nn \\
%%%%%%%%%%%%%%%%%%%%%%%%%%
+ & \frac{\left(2 a-\epsilon _1-\epsilon _2-2 \lambda _1-2 \lambda _2\right) \left(2 a-\epsilon _1-\epsilon _2+2 \lambda _1-2 \lambda _2\right) \left(2 a-\epsilon _1-\epsilon _2+2 \lambda _3-2 \lambda _4\right) \left(2 a-\epsilon _1-\epsilon _2+2 \lambda _3+2 \lambda _4\right)}{32 \left(2 a-\epsilon _1\right) \epsilon _1 \left(2 a-2 \epsilon _1-\epsilon _2\right) \left(\epsilon _1-\epsilon _2\right)} \nn \\
%%%%%%%%%%%%%%%%%%%%%%%%%%
- & \frac{\left(2 a-\epsilon _1-\epsilon _2-2 \lambda _1-2 \lambda _2\right) \left(2 a-\epsilon _1-\epsilon _2+2 \lambda _1-2 \lambda _2\right) \left(2 a-\epsilon _1-\epsilon _2+2 \lambda _3-2 \lambda _4\right) \left(2 a-\epsilon _1-\epsilon _2+2 \lambda _3+2 \lambda _4\right)}{32 \left(2 a-\epsilon _1-2 \epsilon _2\right) \left(2 a-\epsilon _2\right) \left(\epsilon _1-\epsilon _2\right) \epsilon _2} \nn \\
%%%%%%%%%%%%%%%%%%%%%%%%%%
+ & \frac{\left(2 a-\epsilon _1-\epsilon _2-2 \lambda _1-2 \lambda _2\right) \left(2 a-\epsilon _1-\epsilon _2+2 \lambda _1-2 \lambda _2\right) \left(2 a-\epsilon _1-\epsilon _2+2 \lambda _3-2 \lambda _4\right) \left(2 a-\epsilon _1-\epsilon _2+2 \lambda _3+2 \lambda _4\right)}{16 \left(2 a-\epsilon _1\right) \left(2 a+\epsilon _1\right) \left(2 a-\epsilon _2\right) \left(2 a+\epsilon _2\right)} 
\end{align}

\normalsize

We found that 
\be
F_1^+ - \prescript{\diamondsuit}{1}Z^{(0,1)}_{(0,0),(0,0)} = F_1^- - \prescript{\diamondsuit}{1}Z_{(0,1)}^{(0,0),(0,0)} =  \left( {Q\over 2}+\lambda_2\right)\left( {Q\over 2}-\lambda_3 \right)-{3\over 8}+\left(\Delta_1-\Delta_2+\Delta_3-\Delta_4\right)
\ee

The last term exactly compensates for the difference between $F_1^{\pm}$ and ${\cal F}_1^{\pm}$ and this confirms the proposition in the main text on the first level.

\subsection{Second level}

\be
\text{2nd level:} \quad |2\r_{\pm} = t_1^{\pm} L_{-1}^2|0\r_{\pm} + t_2^{\pm} L_{-2}|0\r_{\pm} +
t_3^{\pm} G_{-2}|0\r_{\mp} +  t_4^{\pm} L_{-1}G_{-1}|0\r_{\mp}
\ee

On this level we checked that the following relation holds

\be
\prescript{\diamondsuit}{2}Z^{(0,1)}_{(0,0),(0,0)}=F_2^{+} - K F_1^{+} +{K(K+1) \over 2}
\ee
where

\be
K=   \left( {Q\over 2}+\lambda_2\right)\left( {Q\over 2}-\lambda_3 \right)-{3\over 8}+(\Delta_1-\Delta_2+\Delta_3-\Delta_4)
\ee

\section{Comparison of $\l RRRR \r$ superconformal block with partition functions on the first two levels}

We found on a few first levels
\begin{align}
 |{1\over 2};\pm\r &= \pm { e^{i {\pi \over 4}} \over 2\Delta } (\beta_1 \mp \beta_2)G_{-{1\over 2}} |0;\pm\r \\
		  |1;\pm\r &= {\Delta + \Delta_1 - \Delta_2 \over 2 \Delta } L_{-1} |0;\pm\r \\
 |{3\over 2};\pm\r &= \left( x_1^{\pm}G_{-{3\over 2}}+x_2^{\pm}L_{-1}G_{-{1\over 2}} \right)|0;\pm\r \\
		  |2;\pm\r &= \left( y_1^{\pm}L_{-2}+y_2^{\pm}L_{-1}^2+y_3^{\pm}G_{-{3\over 2}}G_{-{1\over 2}} \right)|0;\pm\r 
\end{align}
where

\begin{align}
x_1^{\pm}(\beta_1,\beta_2) =& e^{i {\pi\over 4}} \frac{\text{$\beta $}_2 (1+2 \Delta +4 \text{$\Delta $}+1-4 \text{$\Delta $}_2) \pm \text{$\beta $}_1 (1+2 \Delta -4 \text{$\Delta $}_1+4 \text{$\Delta $}_2)}{2 (c+2 c \Delta +2 \Delta  (-3+2 \Delta ))} \\
x_2^{\pm}(\beta_1,\beta_2) =& \pm e^{i {\pi\over 4}} \frac{(2\Delta+c)(\beta_1 \mp \beta_2)(1+2\Delta+2\Delta_1-2\Delta_2)-4\Delta(3\beta_1 \mp \beta_2)}{4 \Delta  (c+2 c \Delta +2 \Delta  (-3+2 \Delta ))} \\
y_1^{\pm}(\beta_1,\beta_2) =& -\frac{ (-1+c+8 \text{$\beta $}_1 (-\text{$\beta $}_1 \pm \text{$\beta $}_2)+4 \Delta )+8 e^{i{3\pi \over 4}}  (\pm \text{$\beta $}_1+\text{$\beta $}_2) \left((1-c-2 \Delta ) x_{1}^{\pm}+2 \Delta  x_{2}^{\pm}\right)}{4 (-3+3 c+16 \Delta )} \\
y_2^{\pm}(\beta_1,\beta_2) =& \frac{12 \text{$\beta $}_1 (\text{$\beta $}_1 \mp \text{$\beta $}_2)+16 \Delta ^2+\Delta  (-1+3 c+16 \text{$\Delta $}_1-16 \text{$\Delta $}_2)+3 (-1+c) (\text{$\Delta $}_1-\text{$\Delta $}_2)}{8 \Delta  (-3+3 c+16 \Delta )} + \nn  \\
& + \frac{4 e^{i{3\pi \over 4}} (\pm \text{$\beta $}_1+\text{$\beta $}_2) \Delta  \left(10 x_1^{\pm}+(3+3 c+16 \Delta ) x_{2}^{\pm}\right)}{8 \Delta  (-3+3 c+16 \Delta )}   \\
y_3^{\pm}(\beta_1,\beta_2) =& -\frac{6 \text{$\beta $}_1 (\text{$\beta $}_1 \mp \text{$\beta $}_2)+\Delta +4 e^{i{3\pi \over 4}} (\pm \text{$\beta $}_1+\text{$\beta $}_2) \Delta  \left(5 x_1^{\pm}+3 x_2^{\pm}\right)}{4 \Delta  (-3+3 c+16 \Delta )} 
\end{align}
where the dependence on parameters of external dimensions is shown explicitly for later reference. 
Let us define 
\begin{align}
F_{\pm N} &= {1\over \Gamma_{\pm} \Gamma_{\pm}} \l N;\pm|N\pm \r \\
H_{\pm N} &= {1\over \Gamma_{\pm} \Gamma_{\mp}}\l N;\pm|N;\mp\ \r
\end{align}

We evaluated on the first two levels

\begin{align}
	 F_{\pm{1\over 2}} &= \pm {( \beta_1 \mp \beta_2)(\beta_3 \mp \beta_4) \over 2 \Delta} \\
	 H_{\pm{1\over 2}} &= \pm {( \beta_1 \mp \beta_2)(\beta_3 \pm \beta_4) \over 2 \Delta} \\
	 F_{\pm 1} &= H_{\pm 1} = \frac{(\Delta -\Delta_1+\Delta_2)(\Delta +\Delta_3-\Delta_4)}{2 \Delta } \\
	 F_{\pm{3\over 2}} &= (2\Delta+c) x_1^{\pm}(\beta_2,\beta_1)x_1^{\pm}(\beta_3,\beta_4)+ 4\Delta \Big(x_1^{\pm}(\beta_2,\beta_1)x_2^{\pm}(\beta_3,\beta_4)+ x_2^{\pm}(\beta_2,\beta_1)x_1^{\pm}(\beta_3,\beta_4)\Big) + \nn \\
	 & + 2\Delta(2\Delta+1)x_2^{\pm}(\beta_2,\beta_1)x_2^{\pm}(\beta_3,\beta_4) 	\\
	 H_{\pm{3\over 2}} &= (2\Delta+c) x_1^{\pm}(\beta_2,\beta_1)x_1^{\mp}(\beta_3,\beta_4)+ 4\Delta \Big(x_1^{\pm}(\beta_2,\beta_1)x_2^{\mp}(\beta_3,\beta_4)+ x_2^{\pm}(\beta_2,\beta_1)x_1^{\mp}(\beta_3,\beta_4)\Big) + \nn \\
	 & + 2\Delta(2\Delta+1)x_2^{\pm}(\beta_2,\beta_1)x_2^{\mp}(\beta_3,\beta_4)  \\
	 F_{\pm 2} &= y_1^{\pm}(\beta_2,\beta_1)\left( (4\Delta +{3\over 4}c)y_1^{\pm}(\beta_3,\beta_4) + 6\Delta y_2^{\pm}(\beta_3,\beta_4) + 5\Delta y_2^{\pm}(\beta_3,\beta_4) \right) + \nn \\
	 &+ y_2^{\pm}(\beta_2,\beta_1) \left( 6\Delta y_1^{\pm}(\beta_3,\beta_4) + 4\Delta(2\Delta+1) y_2^{\pm}(\beta_3,\beta_4) + 4\Delta y_2^{\pm}(\beta_3,\beta_4) \right) + \nn \\
	 &+ y_3^{\pm}(\beta_2,\beta_1) \left( 5\Delta y_1^{\pm}(\beta_3,\beta_4) + 4\Delta y_2^{\pm}(\beta_3,\beta_4) + 2\Delta(2\Delta+c+1) y_2^{\pm}(\beta_3,\beta_4) \right)
\end{align}
We present here formulas for conformal blocks on the levels ${3\over 2}$ and $2$ only through the coefficients $x,y$ because explicit formulas are too big.

Then on the first level we found that

\begin{align}
\prescript{\diamondsuit}{1}Z^{(0,0)}_{(1,0)(1,0)} &= H_{-1} + {1\over 2} H_{- {1\over 2}} - U + {1\over 4} \\
\prescript{\diamondsuit}{1}Z^{(0,0)}_{(0,1)(0,1)} &= H_{+1} + {1\over 2} H_{+ {1\over 2}} - U + {1\over 4} \\
\prescript{\diamondsuit}{1}Z^{(0,0)}_{(1,0)(0,1)} &= F_{-1} + {1\over 2} F_{- {1\over 2}} - U + {1\over 4} \\
\prescript{\diamondsuit}{1}Z^{(0,0)}_{(0,1)(0,1)} &= F_{+1} + {1\over 2} F_{+ {1\over 2}} - U + {1\over 4}
\end{align}

And on the second level

\begin{align}
\prescript{\diamondsuit}{2}Z^{(0,0)}_{(1,0)(1,0)} &= H_{-2} + {1\over 2}H_{-{3\over 2}} - U( H_{-1} + {1\over 2} H_{- {1\over 2}})+{1\over 4}H_{-1} +{1\over 4} H_{-{1\over 2}} +
{U(U-1) \over 2} - {U\over 4} + {11\over 64} \\
\prescript{\diamondsuit}{2}Z^{(0,0)}_{(0,1)(0,1)} &= H_{+2} + {1\over 2}H_{+{3\over 2}} - U( H_{+1} + {1\over 2} H_{+ {1\over 2}})+{1\over 4}H_{+1} +{1\over 4} H_{+{1\over 2}} +
{U(U-1) \over 2} - {U\over 4} + {11\over 64} \\
\prescript{\diamondsuit}{2}Z^{(0,0)}_{(1,0)(0,1)} &= F_{-2} - {1\over 2}F_{-{3\over 2}} - U( F_{-1} + {1\over 2} F_{- {1\over 2}})+{1\over 4}F_{-1} +{1\over 4} F_{-{1\over 2}} +
{U(U-1) \over 2} - {U\over 4} + {11\over 64} \\
\prescript{\diamondsuit}{2}Z^{(0,0)}_{(0,1)(1,0)} &= F_{+2} - {1\over 2}F_{+{3\over 2}} - U( F_{+1} + {1\over 2} F_{+ {1\over 2}})+{1\over 4}F_{+1} +{1\over 4} F_{+{1\over 2}} +
{U(U-1) \over 2} - {U\over 4} + {11\over 64}
\end{align}
and similar formulas for $\prescript{\vardiamond}{}Z^{(1,1)}_{(u_1,u_2)(v_1,v_2)}$. The left lower index here denotes the series expansion coefficients of partition function (omitting all other indices ): $Z=\sum  \prescript{}{N}Z \cdot q^N$.
These formulae prove propositions in the main text on the first two levels.

\bibliography{Mylib}{}

\end{document}